\documentclass{revtex4}
\usepackage{url}
\usepackage{ulem}  
\begin{document}
\input epsf.def   

\input psfig.sty


\title{Supernova Neutrino Detection}

\author{Kate Scholberg\\
Department of Physics, Duke University, Durham NC 27708, USA; \\
email: schol@phy.duke.edu}



\begin{abstract} 
A core-collapse supernova will produce an enormous burst of neutrinos of all flavors in the few-tens-of-MeV range.  Measurement of the flavor, time and energy structure of a nearby core-collapse neutrino burst will yield answers to many physics and astrophysics questions.
The neutrinos left over from past cosmic supernovae are also observable,
and their detection will improve knowledge of core collapse rates and average neutrino emission.
This review describes experimental techniques for detection of core-collapse neutrinos, as well as the sensitivities of current and future detectors.
\end{abstract}

\maketitle

\section{INTRODUCTION}

When a massive star has exhausted its nuclear fuel, it collapses to form a compact object such as a neutron star or a black hole.  A prominent feature of the collapse is that $\sim$99\% of the gravitational binding energy of the resulting remnant is converted to neutrinos with energies of a few tens of MeV over a timescale of a few tens of seconds.   This highly efficient energy loss via neutrinos occurs because the neutrinos interact only via the weak interaction and can escape easily, whereas photons are trapped.  
 
Neutrinos were observed for the celebrated 1987A core-collapse supernova (SN1987A) in the Large Magellanic Cloud (LMC), 50~kpc away from Earth.  Two water Cherenkov detectors, Kamiokande-II~\cite{Hirata:1987hu} and the Irvine-Michigan-Brookhaven (IMB) experiment~\cite{Bionta:1987qt}, observed 19 neutrino interaction events between them over a 13-s interval at a time consistent with the estimated time of core collapse.
Two scintillator detectors, 
Baksan~\cite{Alekseev:1987ej} and LSD~\cite{Aglietta:1987it}, also reported observations; the latter report was controversial because the events were recorded several hours early.  The SN1987A events
were certainly nearly all of $\bar{\nu}_e$ flavor.
Although these events were a meager sample, the SN1987A neutrino events were sufficient to confirm the baseline model of core collapse.  Beyond that, they have provided a very wide range of constraints on astrophysics and physics (e.g., References~\cite{Schramm:1990pf,Koshiba:1992yb,Raffelt:1999tx}), resulting in the publication of hundreds of papers, which continues to this day.

Worldwide capabilities for supernova neutrino detection have increased by orders of magnitude since 1987.  The next observation of a nearby core-collapse supernova will provide a great deal of information for both physics and astrophysics.  
The rate of core-collapse supernovae is estimated to be a few per century~\cite{Tammann:1994ev,Cappellaro:2000ez} in a galaxy such as the Milky Way, so the chance of observing one in the next few decades is not negligible.  The most likely distance of the next core-collapse supernova from Earth is between 12 abd 15~kpc, according to the distribution of possible supernova progenitors in the Milky Way~\cite{Mirizzi:2006xx}.

This review concentrates on experimental aspects of the detection of neutrinos from core-collapse supernovae.
The physics of neutrino
interactions in the tens-of-MeV regime imposes limitations on the quality and type
of information that can be obtained from the neutrinos. 

Section~\ref{signal}  briefly describes the main features of the signal.
Section~\ref{interactions}   describes the neutrino interactions that are relevant for current detectors.  Section~\ref{detectors} describes
generic detection issues and backgrounds and covers different detector technology types as well as specific detector examples.  Section~\ref{pointing} addresses techniques for pointing to the supernova.   Sections~\ref{extragalactic} and~\ref{diffuse}  consider detection of neutrinos originating from beyond the Galaxy.  The final section summarizes future prospects.

\section{THE SUPERNOVA NEUTRINO SIGNAL}\label{signal}

Despite enormous recent progress, much about the physics of core collapse is not well understood.  The neutrino messengers from deep inside the supernova will help us understand many aspects of the supernova mechanism and associated phenomena.  The neutrinos are probably intimately involved with the explosion mechanism; imprinted 
on the flux will be signatures of shock waves, accretion, cooling, possible formation of exotic matter, and further collapse to a black hole, and an improved understanding of supernova nucleosynthesis will result from a detection.
A detailed discussion of the physics of core collapse and associated neutrino signatures is
beyond the scope of this review; some example reviews can be found in
References~\cite{Mezzacappa:2005ju,Janka:2006fh,Raffelt:2007nv,Dighe:2008dq}.   

Some general features of the neutrino signal are described briefly as follows.   At the beginning of collapse, one expects a short  (tens-of-milliseconds), bright neutronization or breakout burst dominated by $\nu_e$ from electron capture: $p+e^-\rightarrow n + \nu_e$.  This burst is followed by an accretion phase,
tens to hundreds of milliseconds long, over which electron flavors dominate; at this stage, there may be a complex energy and time structure that represents effects of the standing accretion shock instability phenomenon.
The next phase is cooling, which lasts a few tens of seconds,
during which the core sheds most of its gravitational binding energy.
During this phase, $\nu\bar{\nu}$ pairs dominate neutrino production, and luminosities and temperatures gradually decrease.  An overall feature of the neutrino flux 
is that luminosity is roughly equally divided among flavors.
Furthermore, the flavors have an expected energy hierarchy (where the robustness varies by model) according to
$\langle E_{\nu_e}\rangle < \langle E_{\bar{\nu}_e}\rangle < \langle E_{\nu_x}\rangle$,
where $\nu_x$ represents $\nu_{\mu},\nu_{\tau},\bar{\nu}_\mu$, or $\bar{\nu}_\tau$ (the $\mu$ and $\tau$ flavors have identical production at the energies involved and can be considered collectively).  These energies are in decreasing order of strength of interaction with matter:  $\nu_e$ have more interactions than $\bar{\nu}_e$, because of the excess of neutrons in the core; in turn, $\bar{\nu}_e$ have more interactions than $\nu_x$, which are restricted to neutral currents (NCs).  
The weaker the interactions are, the denser and deeper  the decoupling of the neutrinos from the star and, therefore, the hotter the temperature of the neutrinos at the surface of last scattering (the neutrinosphere).
Some examples of neutrino flux predictions 
in the literature (given in approximate order of modernity)
can be found in References~\cite{Burrows:1991kf,Totani:1997vj,Raffelt:2003en,Fischer:2008rh,Gava:2009pj,Huedepohl:2009wh}.  

\begin{figure}[htb]
 \centering
\centerline{\psfig{figure=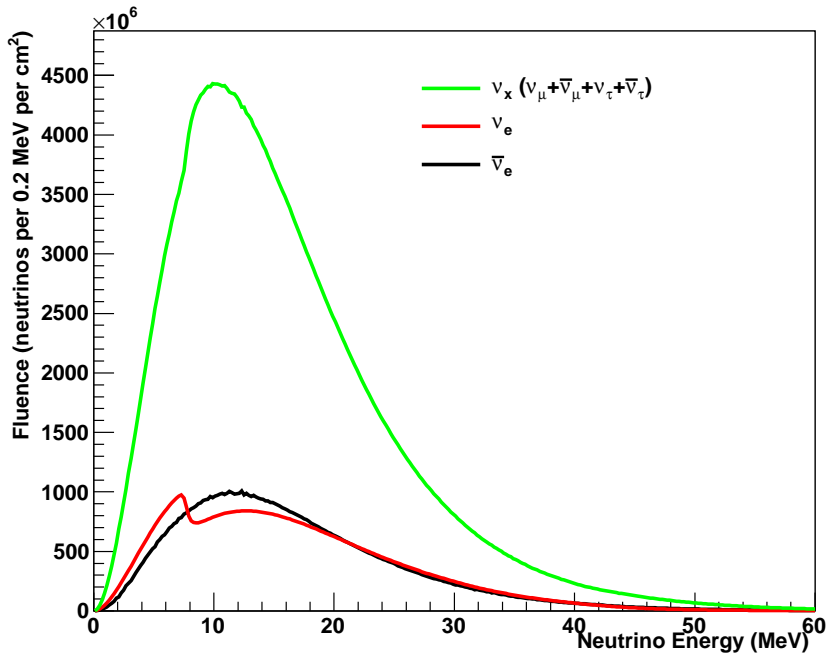,height=4in}}
\caption{Example of expected supernova neutrino spectra (integrated over 10~s) for the different flavor components.  This example is based on the GKVM (Gava-Kneller-Volpe-McLaughlin) model of Reference~\cite{Gava:2009pj}. The prediction includes collective effects, which are responsible for the structure observed in the $\nu_e$ flux.}
\label{figure1}
\end{figure}

Figure~\ref{figure1} shows an example of a flux prediction. Modeling has steadily improved over the past few decades, with inclusion of more and more effects as techniques become more sophisticated and as available computing power increases.
Some of the more recent models tend to produce somewhat cooler spectra and therefore fewer interaction events than older ones.
There may be significant variations in the expected flux 
from supernova to supernova due to differences in the mass and composition of the progenitor, and possibly asymmetries, rotational effects, or magnetic field effects.

Fluxes are expected to have the approximate spectral form
parameterized by (e.g., References~\cite{Huedepohl:2009wh} and ~\cite{Vaananen:2011bf})
\begin{equation}
	\label{eq:PLdistr}
	\phi(E_{\nu}) =N 
	\left(\frac{E_{\nu}}{\langle E_{\nu} \rangle}\right)^{\alpha} \exp\left[-\left(\alpha + 1\right)\frac{E_{\nu}}{\langle E_{\nu} \rangle}\right] \ ,
\end{equation}
where $\alpha$ is often referred to as the ``pinching parameter'', because it controls the high-energy tail of the distribution; $E_\nu$ is the neutrino energy; $\langle E_\nu \rangle$ is the mean neutrino energy; and $N = \frac{(\alpha+1)^{\alpha+1}}{\langle E_{\nu}\rangle\Gamma(\alpha+1)}$is the normalization constant, where $\Gamma$ is 
the Gamma 
function.  
The different $\nu_e$, $\bar{\nu}_e$ and $\nu_x$  flavor components are generally expected to have different $\langle E_\nu \rangle$ and $\alpha$ parameters.  The spectra are expected to evolve in time (e.g., Reference~\cite{Huedepohl:2009wh}).

An enormous pulse of neutrinos from a core collapse will help us understand the physics of neutrinos as well as the astrophysics of core collapse. 
Interesting signatures will manifest themselves as
modulations of the observed spectra, potentially with significant time dependence.
We now know that neutrinos have mass and that the three flavors of neutrinos mix with one another; the supernova neutrino fluxes will also depend on neutrino oscillation parameters.
Matter oscillation (Mikheyev-Smirnov-Wolfenstein) effects modify the spectra as the neutrinos traverse dense matter, so the signals will depend both on neutrino oscillation parameters and on supernova matter profiles.
In fact, conditions in the supernova will be so extreme that neutrino-neutrino interactions will become important, which will lead to a complex phenomenology of collective effects.  These effects include spectral swap or spectral split features in the spectra, for which
flavors are effectively swapped above or below a particular energy threshold.
The recent literature exploring supernova collective oscillation  physics is extensive: A non-comprehensive sample can be found in References~\cite{Duan:2005cp,Fogli:2007bk,Raffelt:2007cb,Raffelt:2007xt,EstebanPretel:2008ni,Duan:2009cd,Dasgupta:2009mg,Duan:2010bg,Duan:2010bf}. 

The nature of the neutrino spectra and their time evolution depend on mass and  oscillation parameters, such as $\theta_{13}$ and the 
mass hierarchy.  For the normal hierarchy, $m_3 >> m_2,m_1$, there are two light and one heavy neutrino mass states, and for the inverted hierarchy, $m_1,m_2 >> m_3$, there are two heavy and one light neutrino mass states.
In particular, the spectra of $\nu_e$ and $\bar{\nu}_e$ will differ according to the mass hierarchy, so supernova neutrino measurements could tell us what the hierarchy is (e.g., References~\cite{Dighe:1999bi,Dighe:2003be,Lunardini:2003eh,Dighe:2008dq, Chakraborty:2011ir,Choubey:2010up}).
Note that although reactor and beam experiments have now measured the value of $\theta_{13}$, the mass hierarchy may not be determined by laboratory experiments for a decade or more~\cite{Akiri:2011dv}.
Furthermore,  the chance that the supernova neutrinos will traverse Earth matter on their way to a detector is not negligible~\cite{Mirizzi:2006xx} and
oscillations in the Earth modulate the supernova neutrino spectrum for either $\nu_e$ or $\bar{\nu}_e$~\cite{Lunardini:2001pb,Takahashi:2001dc,Takahashi:2002cm} depending on the hierarchy.  Therefore,
information on oscillation parameters can be determined in a supernova-model-independent way if measurements at different locations on the Earth can be compared.
Even in a single detector, an Earth-matter-induced spectral modulation may give information about oscillations~\cite{Dighe:2003be,Dighe:2003jg,Dighe:2003vm}.
Furthermore the flux may carry signatures of oscillations involving sterile neutrino states (e.g.,~\cite{Fetter:2002xx}).

Although they probably will not be competitive with near-future laboratory measurements (or cosmological measurements), constraints on neutrino absolute mass may also become possible via measurements of an energy-dependent time delay, especially if some sharp feature in the spectrum is available to set a time reference (e.g., References~\cite{Beacom:1998ya,Beacom:2000qy}).  Other neutrino properties, such as lifetime, magnetic moments, nonstandard interactions,
and the speed from time of flight over a known distance, can also be constrained~\cite{Schramm:1990pf,Raffelt:1999tx}.
The physics to be learned is not confined to the neutrino sector.  There will be opportunities to shed light on other particle physics as well~\cite{Raffelt:1999tx}, given that the neutrino signal will provide a measure of the total energy of the collapse and constrains energy loss
via new particles and exotic mechanisms, including axions~\cite{Raffelt:1999tx} and extra dimensions~\cite{Hannestad:2001jv}.

A generic potential difficulty is that both core-collapse physics and neutrino
physics  affect the nature of the neutrino burst, and it may not be trivial to
disentangle the two.  Relatively robust and model-independent signatures 
do exist, however.  Clearly, the more experimental
data we can gather about the flavor, energy and time structure
of the burst, in as many detectors around the world as possible,
the better our chances will be of disentangling the various effects.

\section{NEUTRINO INTERACTIONS IN THE TENS-OF-MEV RANGE}\label{interactions}

Neutrinos are detected via electromagnetically or strongly interacting  products
of weak charged-current (CC) and NC interactions with electrons and nuclei.   This section describes what is known about neutrino interactions
in the tens-of-MeV range relevant for current detectors, along with their observables.
Unfortunately, relatively few interactions have precisely known cross sections.  Except for elastic scattering and inverse $\beta$ decay (IBD) interactions, both theoretical and experimental knowledge is limited.

\subsection{Inverse Beta Decay}\label{ibd}

Relatively cheap detector materials such as water and hydrocarbon-based scintillator have
many free protons.
The most significant interaction in these materials is
IBD, which is the
interaction between $\bar{\nu}_e$ and free protons, 
$\bar{\nu}_e + p \rightarrow n + e^+$.   The IBD kinematic threshold is $E_{\nu_{\rm thr}}=1.8$~MeV.
The positron's
energy loss can typically be observed.
In the supernova energy regime, to a good approximation $E_{e^+}=E_\nu-1.3$~MeV.  
In scintillation detectors, the 0.511-MeV positron annihilation $\gamma$s may also
be observed.
 The neutron may 
be captured on free protons, with an approximately 200-$\mu$s thermalization and capture time,
producing a deuteron and a 2.2~MeV $\gamma$.
The neutron may also be captured on another nucleus; in particular, the detector may be doped with some material with a high neutron capture cross section, such as gadolinium (Gd).  Gd of natural isotopic composition 
has average neutron capture cross section some $1.6\times 10^5$ times that of free protons, and a thermalization and 
capture time of a few tens of microseconds when dissolved in water or scintillator.  Neutron capture on a Gd nucleus is followed
by a deexcitation cascade of $\gamma$s summing to approximately 8~MeV of energy.
In scintillator, most of the energy is visible via 
Compton scattering of the $\gamma$s.
Reference~\cite{Beacom:2003nk} proposes doping of water with a Gd compound, for which
approximately $4$~MeV of energy is visible per neutron capture~\cite{Watanabe:2008ru}.

The calculated IBD cross section can be found in~\cite{Strumia:2003zx}.
The interaction has only a slight energy-dependent anisotropy~\cite{Vogel:1999zy}, so in general it
is not very useful for pointing to a supernova.

\subsection{Elastic Scattering on Electrons}\label{es}

Although the cross section is relatively small compared with other interactions,
neutrino-electron elastic scattering, $\nu_x+e^-\rightarrow \nu_x+e^-$,  is 
important because of its directionality.  The electron is scattered 
in the direction of the incoming neutrino; thus, for detectors that
can reconstruct electron tracks, such as water Cherenkov and liquid
argon time-projection chambers (TPCs), the elastic scattering sample can be selected against background.  Furthermore, elastic scattering events can be used to point to the supernova
(see Section~\ref{pointing}).
Elastic scattering proceeds via both 
CC and NC interactions for $\nu_e$ and $\bar{\nu}_e$ (the latter are helicity suppressed), and via NC interaction for $\nu_x$.
The cross sections, and the angular and recoil energy distributions of the scattered electrons, are based on very
well understood weak interaction physics~\cite{Marciano:2003eq}.

\subsection{Charged- and Neutral-Current Interactions with Nuclei}

Neutrinos also interact with nucleons in nuclei via CC and NC interactions, although cross sections are typically somewhat smaller for bound than for free nucleons.
CC interactions proceed via interaction of $\nu_e$ and $\bar{\nu}_e$ with neutrons and protons, respectively, in nuclei,
$\nu_e+(N,Z)\rightarrow (N-1,Z+1) +e^-$, and 
$\bar{\nu}_e+(N,Z)\rightarrow (N+1,Z-1) +e^+$.  The kinematic threshold 
is $E_{\nu_{\rm thr}}=\frac{M_f^2+m_e^2+2 M_f m_e - M_i^2}{2M_i}\sim M_f-M_i+m_e$, where $M_f$ and $M_i$ are the initial and final state nuclear masses and $m_e$ is the electron mass.
Note that at supernova energies, $\nu_\mu$ and $\nu_\tau$ are below the CC interaction threshold and thus are kinematically unable to produce their partner leptons. In nuclei, the $\bar{\nu}_e$ interaction is typically suppressed at a given energy 
with respect to the $\nu_e$ interaction due to Pauli blocking.  Typically the energy loss of the charged lepton is observable.  In the $\bar{\nu}_e$ case, the $\gamma$s produced by annihilation of the positron may be observable.  Ejecta (nucleons and $\gamma$s) produced by the final nucleus as it deexcites may also be observable and may help to tag the interaction.
The neutrino interactions may furthermore leave radioactive nuclear products, for which decays correlated in space and time with the primary lepton could be observed~\cite{Kolbe:2002gk}.

NC interactions on nucleons in nuclei may also produce observable signals via ejected nucleons or deexcitation $\gamma$s.  Of particular note are the 15.1-MeV $\gamma$ from $^{12}$C deexcitation and the ejection of neutrons from lead (see Sections~\ref{scint} and~\ref{lead}).

In CC cases, produced leptons retain memory
of the incoming neutrino energy, as the heavy recoil nucleus tends to take away little energy.
In both CC and NC cases, specific interactions have different thresholds and energy-dependent cross sections, so even if there is no product to measure that remembers the neutrino energy on an event-by-event basis, there is neutrino spectral information to be had by measuring statistical distributions of products.

Neutrino interactions on nuclei in the tens-of-MeV range are relatively poorly understood theoretically, in terms of both the interaction rate and the angular and energy distributions of the resulting interaction products.
These interactions and their products tend to be sensitive to the details of the nuclear physics involved.   
Nevertheless, several calculations exist.
Cross sections on carbon are considered in References~\cite{Kolbe:1995af,Kolbe:1999au,Volpe:2000zn}.
Cross sections and signatures for oxygen that are relevant for water detectors are considered in References~\cite{Haxton:1987kc,Haxton:1988mw,Haxton:1990ks,Langanke:1995he,Kolbe:2002gk}.
Calculations relevant for the neutrino response of argon detectors can be found in References~\cite{Raghavan:1986hv,Ormand:1994js,
Bhattacharya:1998hc,
GilBotella:2003sz,
Bueno:2003ei,
GilBotella:2004bv,
SajjadAthar:2004yf},
and studies relevant for lead and iron appear in
References~\cite{Fuller:1998kb,
Kolbe:2000np,
Volpe:2001gy, 
Toivanen:2001re, Engel:2002hg,
Samana:2008pt}.

Uncertainties in theoretical calculations are no better than the $\sim$10-20\% level,  and unfortunately experimental measurements of neutrino interactions in this energy range are poor or nonexistent.  The only measurements with $\sim$10-20\% uncertainties are for $^{12}$C by the Liquid Scintillator Neutrino Detector (LSND) at Los Alamos National Laboratory and KARMEN at Rutherford Appleton Laboratory~\cite{Armbruster:1998gk,Auerbach:2001hz}.
Future programs to measure cross sections will be highly desirable for an effective interpretation of the next observed supernova burst signal, as well as for the understanding of processes in the supernova itself.  A stopped-pion source (as used for the LSND and KARMEN measurements),
such as the Spallation Neutron Source~\cite{Efremenko:2005nf,SajjadAthar:2005ke} or a cyclotron-based source~\cite{Alonso:2010fs} would be nearly ideal: The well-understood stopped-pion spectra of $\nu_\mu$, $\bar{\nu}_\mu$ and $\nu_e$, with neutrino energies around $\sim$20-50~MeV, are a rather close match to the expected supernova spectrum.  Low-energy $\beta$ beams, neutrinos originating from the decay of radioactive ion beams, with tunable spectra, are also promising sources of well-understood flux
 for measurements of supernova-relevant cross sections~\cite{Serreau:2004kx,Jachowicz:2008kx}.

\subsection{Coherent Elastic Neutrino-Nucleus Scattering}

Neutrinos at MeV energies will also coherently scatter off protons or entire nuclei via NC weak interactions.  The interaction rate is relatively 
high;
 however recoil energies tend to be rather low, so detection thresholds need to be very low.  For coherent scattering on protons, recoil energies are less than a few MeV but within the reach of low-background
large scintillation detectors~\cite{Beacom:2002hs}.
When entire nuclei are the target, recoil energies tend to be in the few-keV to few-tens-of-keV range~\cite{Freedman:1977xn,Drukier:1983gj}, out of reach of conventional detectors.  However, detectors designed to search for 
weakly interacting massive particle (WIMP) dark matter recoils will be sensitive~\cite{Horowitz:2003cz}.

\subsection{Summary of Interactions}
\begin{figure}[h]
\centerline{\psfig{figure=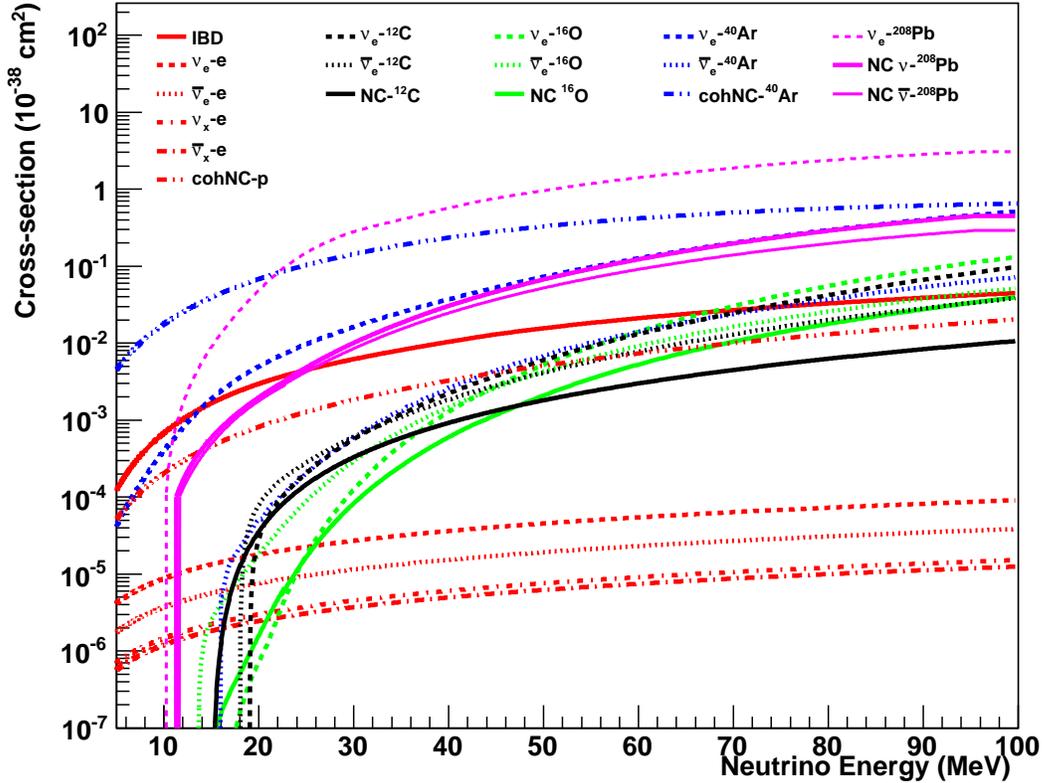,height=5in}}
\caption{Cross sections per target for relevant interactions: see \texttt{http://www.phy.duke.edu/$\sim$schol/snowglobes}  for references for each cross section plotted.}
\label{figure2}
\end{figure}

Table~\ref{tab:interactions} summarizes interactions on different targets relevant for existing detector types and lists the primary detection modes.  Estimates of the number of interactions per kt (before considering the detector response) are given in the table for two different models from the literature, the Livermore model~\cite{Totani:1997vj} and the GKVM (Gava-Kneller-Volpe-McLaughlin) model~\cite{Gava:2009pj}.  Note that there can be considerable variation in total neutrino interaction yield from model to model, depending on temperature and pinching, because of the energy dependence of the cross sections. 
Figure~\ref{figure2} shows cross sections for relevant processes as a function of neutrino energy. See \texttt{http://www.phy.duke.edu/$\sim$schol/snowglobes} for complete references and assumptions.

\begin{table}[h]
\def~{\hphantom{0}}
\caption{Summary of relevant interactions for current and near-future detectors.   The observables column lists primary observable products relevant for interactions in current detectors.  Abbreviations: C,  energy loss of a charged particle; N,  produced neutrons; G, deexcitation $\gamma$s; A, positron annihilation $\gamma$s.
Note there may in principle be other signatures for future detector technologies or detector upgrades.  
The interactions column gives interactions per kt at 10 kpc for two different neutrino flux models (Livermore/GKVM), for neutrino energies greater than 5 MeV, computed according to 
\texttt{http://www.phy.duke.edu/$\sim$schol/snowglobes}.  No detector response is taken into account here, and actual detected events may be significantly fewer.   For elastic scattering and inverse $\beta$ decay, the numbers per kt refer to water; for other detector materials, the numbers need to be scaled by the relative fraction of electrons or protons, respectively.  For neutrino-proton elastic scattering, the numbers per kt refer to scintillators.
\label{tab:interactions}}
\begin{tabular}{@{}|l|l|l|@{}}%
\hline 
Channel &  Observable(s) & Interactions  \\ \hline
$\nu_x + e^- \rightarrow \nu_x + e^-$            &    C   & 17/10 \\ 
$\bar{\nu}_e+ p \rightarrow e^+ + n$           &  C, N, A    &  278/165  \\ 
$\nu_x+ p \rightarrow  \nu_x + p$           & C   &  682/351 \\

$\nu_e + {}^{12}{\rm C} \rightarrow e^- + {}^{12}{\rm N^{(*)}}$    &   C, N, G               &  3/9 \\

$\bar{\nu}_e + {}^{12}{\rm C} \rightarrow e^+ + {}^{12}{\rm B^{(*)}}$  & C, N, G, A&  6/8  \\

$\nu_x + {}^{12}{\rm C} \rightarrow \nu_x+ {}^{12}{\rm C}^*$  & G &  68/25 \\ 

$\nu_e + {}^{16}{\rm O} \rightarrow e^- + {}^{16}{\rm F^{(*)}}$     &  C,  N, G              & 1/4  \\
$\bar{\nu}_e + {}^{16}{\rm O} \rightarrow e^+ + {}^{16}{\rm N^{(*)}}$  & C, N, G & 7/5  \\ 

$\nu_x + {}^{16}{\rm O} \rightarrow \nu_x+ {}^{16}{\rm O}^*$  &   G           &  50/12 \\  

$\nu_e + {}^{40}{\rm Ar} \rightarrow e^- + {}^{40}{\rm K}^*$    &    C, G                 &  67/83 \\

$\bar{\nu}_e + {}^{40}{\rm Ar} \rightarrow e^+ + {}^{40}{\rm Cl}^*$  & C, A, G & 5/4  \\


$\nu_e + {}^{208}{\rm Pb} \rightarrow e^- + {}^{208}{\rm Bi}^*$    & N                  &  144/228 \\ 
$\nu_x + {}^{208}{\rm Pb} \rightarrow \nu_x+ {}^{208}{\rm Pb}^*$  &   N            &  150/55 \\  

$\nu_x + A \rightarrow \nu_x + A$  & C & 9,408/4,974\\

\hline
\end{tabular}
\end{table}

\section{TECHNOLOGIES FOR SUPERNOVA NEUTRINO DETECTION}\label{detectors}

The particles produced by supernova neutrinos are detected via numerous standard particle-detection techniques, namely by collecting photons or charge proportional to ionization energy loss, collecting Cherenkov photons, or detecting neutrons.  A general review of relevant detection physics can be found in Reference~\cite{Nakamura:2010zzi}.  
The specific experimental techniques employed depend on the characteristics of the target; in many cases, the target material itself is the detection medium. 

A neutrino experimentalist's  basic strategy is
to collect as many neutrino events from a supernova burst 
as possible, of as many
flavors as possible.  For current detectors, typical event yields are a few hundred
events per kt of
detector material for a core-collapse event just beyond the center of the Milky Way, 10~kpc away. 
Desirable for measurement are event-by-event timing resolution, the ability to
measure neutrino energies with good resolution, and the ability to use the
neutrino information to point back to the supernova.  Sensitivity to
all flavors of the burst is especially desirable: $\nu_\mu$ and
$\nu_\tau$ flavors constitute two thirds of the burst's luminosity, but NC
sensitivity is required to detect them.   Note that the interaction rate is not
the only thing that matters: It is 
especially valuable for detectors to have the ability to tag
interactions as $\nu_e$, $\bar{\nu}_e$, and $\nu_{x}$, in addition to simply collecting them.  An ideal detector would measure the flavor, energy, time, and direction of the neutrinos on an event-by-event basis (with no background), but 
in reality, one must settle for imperfectly reconstructed events and inferred statistical information.

The expected number of events from a supernova should scale simply with distance as 1/$D^2$, where $D$ is the distance to the supernova.  To a good approximation for most technologies (to the extent that detection efficiency is independent of detector size), event rates scale linearly with detector mass.  

Expected differential event rates of observed particles for a given neutrino interaction process for a realistic detector can be computed by folding a given supernova neutrino flux with the cross section and detector response according to:
 
\begin{equation}
\frac{dn}{dE'} = \int_0^\infty \int_0^\infty dE d\hat{E}\Phi(E) \sigma(E) k(E-\hat{E}) T(\hat{E}) V(\hat{E}-E') 
\end{equation}

where $E$ is the neutrino energy;
$\hat{E}$ is the produced particle energy; $E'$ is the measured product particle energy, 
$\sigma(E)$ is the total cross section of the process, $k(E-\hat{E})$ gives the energy distribution of the produced particle; and 
$T(\hat{E})$ and $V(\hat{E}-E')$ model the detector threshold and energy response, respectively.
Here we employ \textsf{SNOwGLoBES} (\texttt{http://www.phy.duke.edu/$\sim$schol/snowglobes}), which makes use of  \textsf{GLoBES} (\texttt{http://www.mpi-hd.mpg.de/lin/globes}) to compute the event rates.

For a successful supernova burst observation, the detector
background rate must not greatly exceed the signal rate in a $\sim$10-s burst.  Backgrounds for supernova neutrino detection vary by detector type and location.  
Ambient radioactivity, from the environment or detector materials (\textit{e.g.} uranium/thorium chain isotopes and radon), is nearly always of concern, although typically such background events do not have energies higher than $\sim$7-10~MeV, and a relatively high fraction of burst signal events exceed that energy threshold.   Nevertheless, radioactivity can be troublesome for measurements of the low-energy end of the signal, possibly at late times,  and for low-threshold detectors (see Section~\ref{lowthresh}).

Cosmic-ray-related backgrounds 
can be suppressed by siting detectors deep underground. However,
cosmic-ray muons can still penetrate to deep sites and produce nuclear fragments via spallation or capture processes with atoms in a detector or surrounding materials.  These spallation fragments include radioactive nuclei and neutrons or other hadrons, which later interact with nuclei; the various products can decay on timescales ranging from less than milliseconds to hours, days or longer.   
The distribution of products is not well understood for all target nuclei.
Muon spallation events can produce fake bursts over timescales of tens of seconds.
In principle, if one can identify the parent muon of a spallation event or require spatial uniformity (because spallation products are correlated with the linear track of a muon), one can reject spallation background, at some cost in dead time; however in no detector can this be done with perfect efficiency.
Although cosmic-ray muons get rarer the deeper one goes, their energy spectrum gets harder and the likelihood per penetrating muon of producing spallation increases.

Other backgrounds include reactor $\bar{\nu}_e$, solar $\nu_e$, and low-energy atmospheric neutrinos and antineutrinos-- however, all of these backgrounds should be very small for Galactic bursts.  In addition, detectors are subject to various kinds of instrumental noise or misreconstruction, the nature of which is highly specific to the particular detector setup (and in practice can vary in time significantly).   
However for current underground detectors, background rates should be very low for the duration of a Galactic supernova burst.  Supernova neutrino detection is conceivable even for some surface detectors~\cite{Sharp:2002as,Ayres:2004js}. 
Backgrounds become a dominant concern for more distant supernovae (see Sections~\ref{extragalactic} and \ref{diffuse}), even for deep detectors.

\subsection{Scintillation Detectors}\label{scint}

Scintillator detectors are composed of hydrocarbons, which have the approximate chemical formula C$_n$H$_{2n}$.
The energy loss of charged particles is observed via light emitted from deexcitation of molecular energy levels, and a very large number of photons may be released.
Large-mass scintillator detectors typically take the form of large homogeneous volumes of
liquid viewed by photomultiplier tubes (PMTs); they may also be segmented into smaller volumes.  The energy loss is proportional to the number of photons collected, and the interaction vertices may be reconstructed through the use of the time-of-arrival information of the photons.
Large numbers of photoelectrons may be collected for typical PMT densities, leading to excellent energy resolution and low thresholds.  

Because of the presence of free protons in scintillator, IBD is hugely dominant for a supernova burst signal.  In scintillator, a neutron produced in an IBD interaction is thermalized and captured on a free proton, $n+p \rightarrow d+ \gamma$, with a $\sim$200-$\mu$s delay;
the energy loss from Compton scattering of the produced 2.2-MeV $\gamma$ is also observable.
Furthermore, the two 0.511-MeV $\gamma$s from the annihilation of the positron  also contribute to the observed energy loss (although fine segmentation is usually required to distinguish them).
Elements with a high neutron capture cross section may be dissolved in the scintillator.  In particular, Gd is highly effective for enhancing neutron capture-detection efficiency: see Section~\ref{ibd}.
The time sequence of a prompt positron signal followed by a delayed capture signal provides a clean IBD tag.

Examples of large homogeneous scintillation detectors are  KamLAND~\cite{Eguchi:2002dm} in Japan, and  Borexino~\cite{Cadonati:2000kq,Monzani:2006jg} in Italy.  Segmented detectors with supernova sensitivity include the past MACRO detector~\cite{Ambrosio:1997hh} and the current detectors LVD~\cite{Aglietta:1992dy,Agafonova:2007hn} in Italy and Baksan~\cite{Alekseev:1993dy} in Russia.  Future detectors include SNO+~\cite{Kraus:2010zzb}, currently under construction at SNOLAB in Canada, and the proposed experiments LENA~\cite{Wurm:2011zn} and HanoHano~\cite{Learned:2008zj}.
Some surface detectors may also have sensitivity; these include 
MiniBooNE~\cite{Sharp:2002as,AguilarArevalo:2009ju}, a kt mineral-oil detector at Fermi National Accelerator Laboratory that makes use primarily of Cherenkov light for particle detection, and NO$\nu$A~\cite{Ayres:2004js}, a segmented liquid scintillator detector.  For such detectors, despite relatively large cosmic-ray rates, a burst from a nearby supernova would stand out above the background.
Smaller surface experiments with dissolved Gd, designed for reactor oscillation studies-- Double Chooz~\cite{Ardellier:2006mn}, RENO~\cite{Ahn:2010vy} and Daya Bay~\cite{OchoaRicoux:2011zz}-- will be able to gather some tens of IBD events at 10~kpc with a good signal-to-background ratio, thanks to excellent tagging.

Elastic scattering will contribute a few percent to the total supernova burst event rate; however, because light emission is isotropic, little supernova direction information will be obtained.  Nevertheless, for very good vertex resolution, some directional information can be observed via separation of neutron capture and positron signal sites for IBD
(see Section~\ref{pointing}).

Neutrino interactions on carbon also occur.  In particular, a 15.1-MeV deexcitation $\gamma$ from NC excitation of $^{12}$C may be observable, and good energy resolution in scintillator would allow determination of the total neutrino flux.

Finally, in scintillator, it may be possible to observe NC neutrino-proton elastic scattering.  This signal produces very low recoil energy protons; in scintillator, quenching effects suppress light from this signal.  Nevertheless, significant numbers of events could be observed~\cite{Beacom:2002hs}, and measurements of the recoil energy distribution is one of the few ways to obtain information about the $\nu_x$ spectra~\cite{Dasgupta:2011wg}.

\begin{figure}[htb]
  
\centerline{\psfig{figure=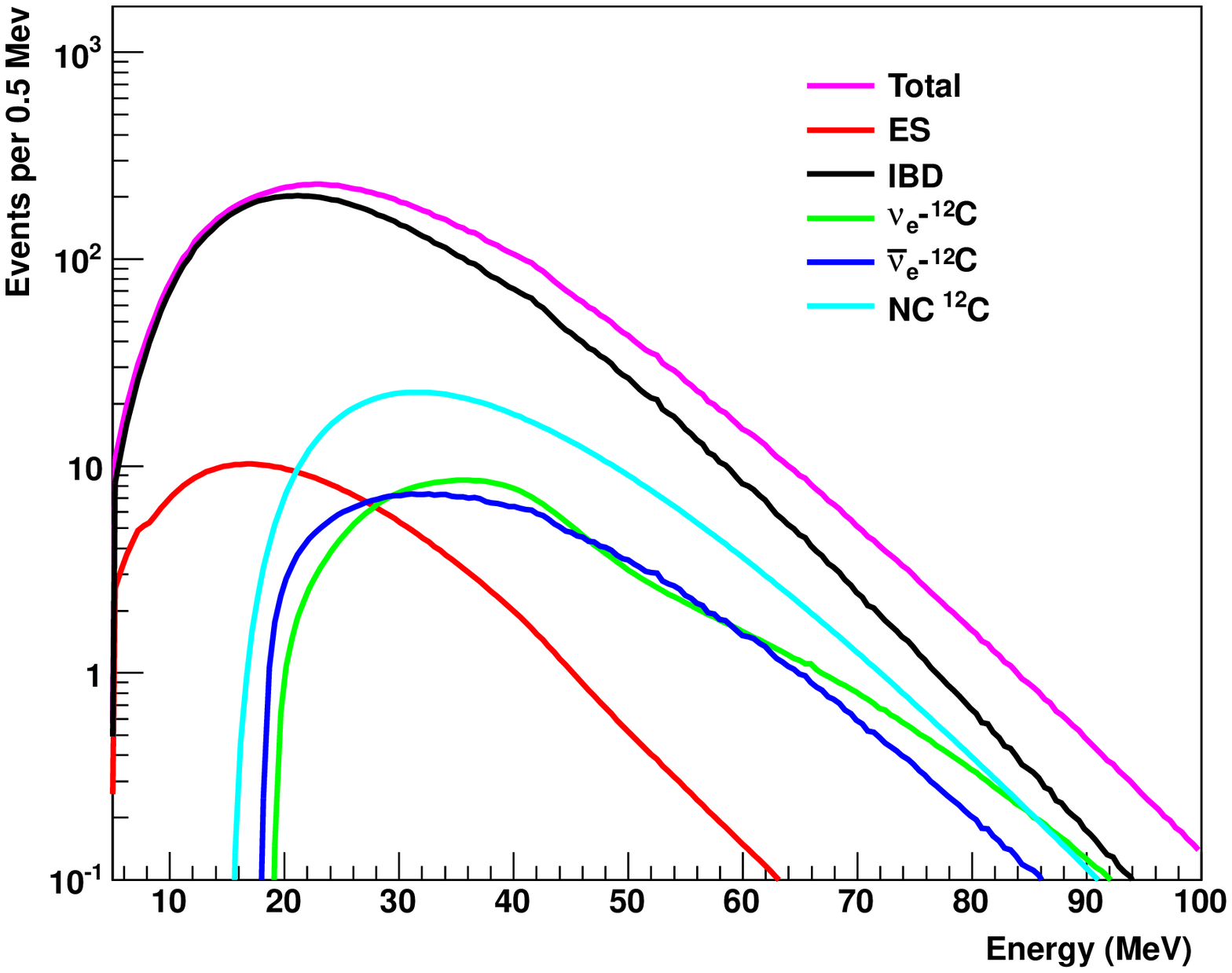,height=3in}}
\centerline{\psfig{figure=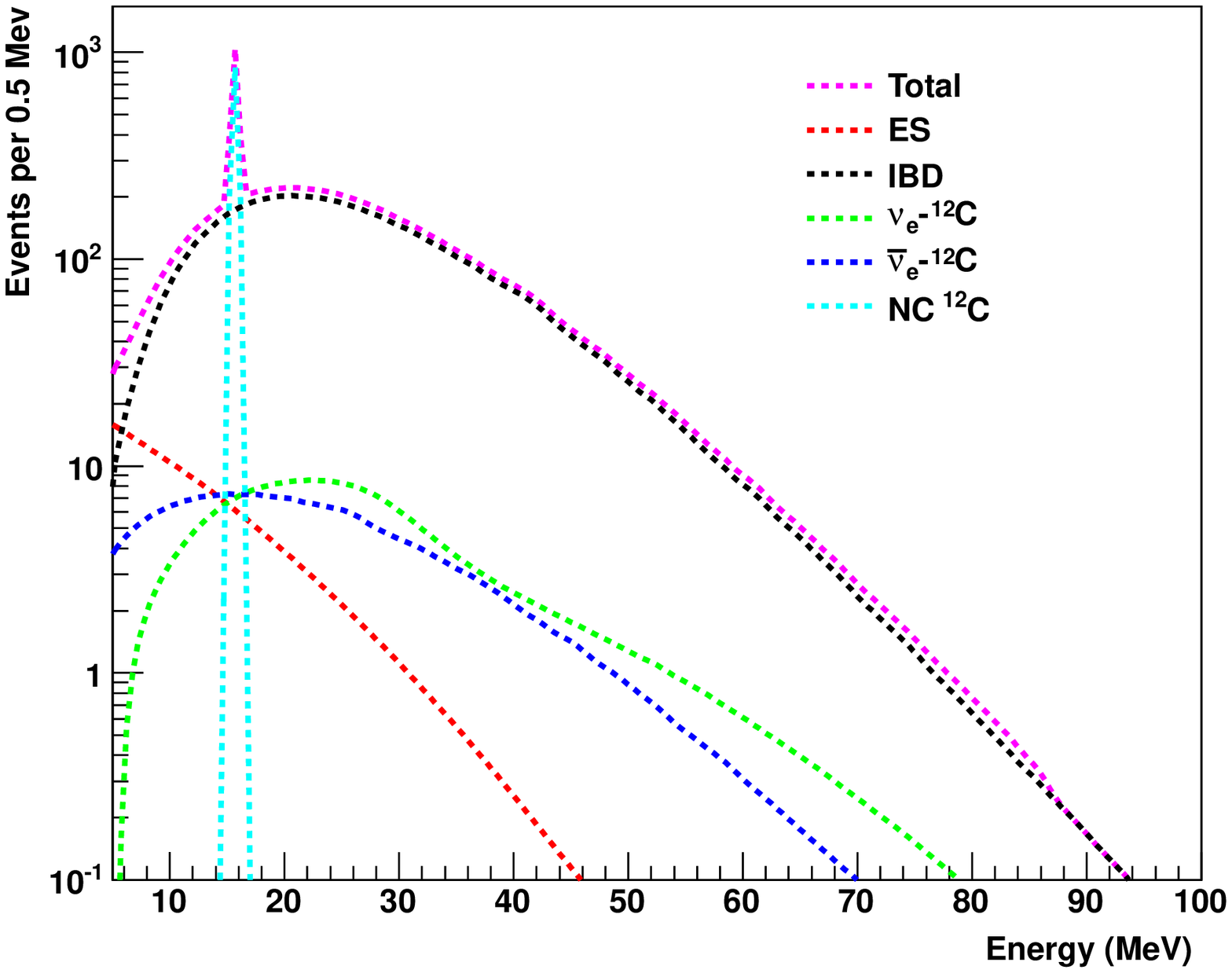,height=3in}}

  \caption{Event rates in 50~kt of scintillator for the GKVM (Gava-Kneller-Volpe-McLaughlin) model 
  (events per 0.5~MeV). Top: interaction rates as a
    function of true neutrino energy.  Bottom:  Smeared
    rates as a function of detected energy (taking into account energy resolution).}
  \label{figure3}
\end{figure}

Figure~\ref{figure3} shows an example of expected event rates in a scintillator detector, computed according to \texttt{http://www.phy.duke.edu/$\sim$schol/snowglobes}  and
assuming an estimated energy resolution 
from Reference~\cite{Eguchi:2002dm}.  The assumed flux is from Reference~\cite{Gava:2009pj}.

\subsection{Water Cherenkov Detectors}\label{wc}

Water is another substance with an abundance of free protons; also, it has the advantage of being inexpensive, so large detectors are possible. Like scintillator detectors, water Cherenkov detectors consist of homogeneous volumes of liquid instrumented with PMTs.  Charged particles are detected via their Cherenkov light emission, which forms a 42$^\circ$ cone for relativistic particles. 
The energy loss is proportional
to the number of photons detected, and one may reconstruct the charged
particle's interaction vertex and direction via the Cherenkov ring
pattern.
Cherenkov detection produces relatively little light compared with that produced by scintillator (by approximately a factor of 50) and has the disadvantage that particles must have speeds exceeding $c/n$ to be visible (here, $n$ is the index of refraction of the material).  Water's index of refraction is $n\sim 1.34$; for this value, the (total-energy) Cherenkov
threshold for electrons is 0.8~MeV, and for protons it is 1,400~MeV.  
As a result, protons induced by supernova neutrinos are always invisible, as are sub-MeV electrons; low energy $\gamma$s, which interact by Compton scattering and for which scattered electrons are often below the Cherenkov threshold, often have rather low light yields and poor detection efficiency.   However, Cherenkov detection  has the advantage that the observed photons carry directional information.  Therefore, for anisotropic interactions such as elastic scattering, angular distributions of products can be measured and used both for supernova pointing and for the disentanglement of flavor components~\cite{Scholberg:2011zzb,Akiri:2011dv}.

As for scintillator, interaction rates in water are dominated by IBD.  Due to the Cherenkov threshold, the detection of the 2.2-MeV $\gamma$ from neutron capture on free protons is very difficult.  It may be possible to dissolve Gd compounds in the water to enhance neutron tagging~\cite{Beacom:2003nk}.  Approximately 4~MeV of equivalent energy can be detected per neutron capture; in Super-Kamiokande (Super-K), the neutron tagging efficiency is estimated to be approximately 67\%~\cite{Watanabe:2008ru}.
A possible Gd enhancement is being studied for Super-K~\cite{Kibayashi:2009ih}.  Neutron tagging will be helpful for determining the total flavor components of the supernova burst signal~\cite{Akiri:2011dv}.

Other interactions in oxygen nuclei of water will contribute to the total burst signal in a water detector.  
Deexcitation $\gamma$s from NC interactions~\cite{Langanke:1995he} may be visible, although due to Cherenkov threshold effects, detection efficiency is poor and requires relatively high PMT coverage~\cite{Scholberg:2011zzb,Akiri:2011dv}.
Some interactions have ejected neutrons~\cite{Kolbe:2002gk}; these may be captured and detected with Gd loading (which may lead to confusion with IBD events).
Angular information obtained from the predicted CC $^{16}$O backwards anisotropy may also be usable~\cite{Haxton:1988mw}.

Past examples of water Cherenkov detectors include IMB~\cite{BeckerSzendy:1992hr} and Kamiokande
~\cite{Hirata:1991ub}, both of which are famous for the SN1987A observation.
The only current example of a large water Cherenkov detector is Super-K~\cite{Ikeda:2007sa} in Japan.  Future concepts include an initially proposed water Cherenkov detector design for the Long-Baseline Neutrino Experiment (LBNE)~\cite{Akiri:2011dv}, Hyper-Kamiokande~\cite{Abe:2011ts}, and
MEMPHYS~\cite{Borne:2011zz}.

Although there are no current or planned future instances of heavy-water detectors, the Sudbury Neutrino Observatory (SNO) detector~\cite{Boger:1999bb} in Canada, which took data between 1999 and 2006, was sensitive to supernova neutrinos
 via NC and CC breakup interactions 
($\nu_e+d\rightarrow p+p+e^-$,
$\bar{\nu}_e+d\rightarrow n+ n +e^+$), in its 1~kt of heavy water. 
These interactions were observable via Cherenkov light and multiple methods of neutron detection, in addition to interactions in SNO's 1.7~kt of light water~\cite{Fleurot:2007zz}.

\begin{figure}[htb]
  
\centerline{\psfig{figure=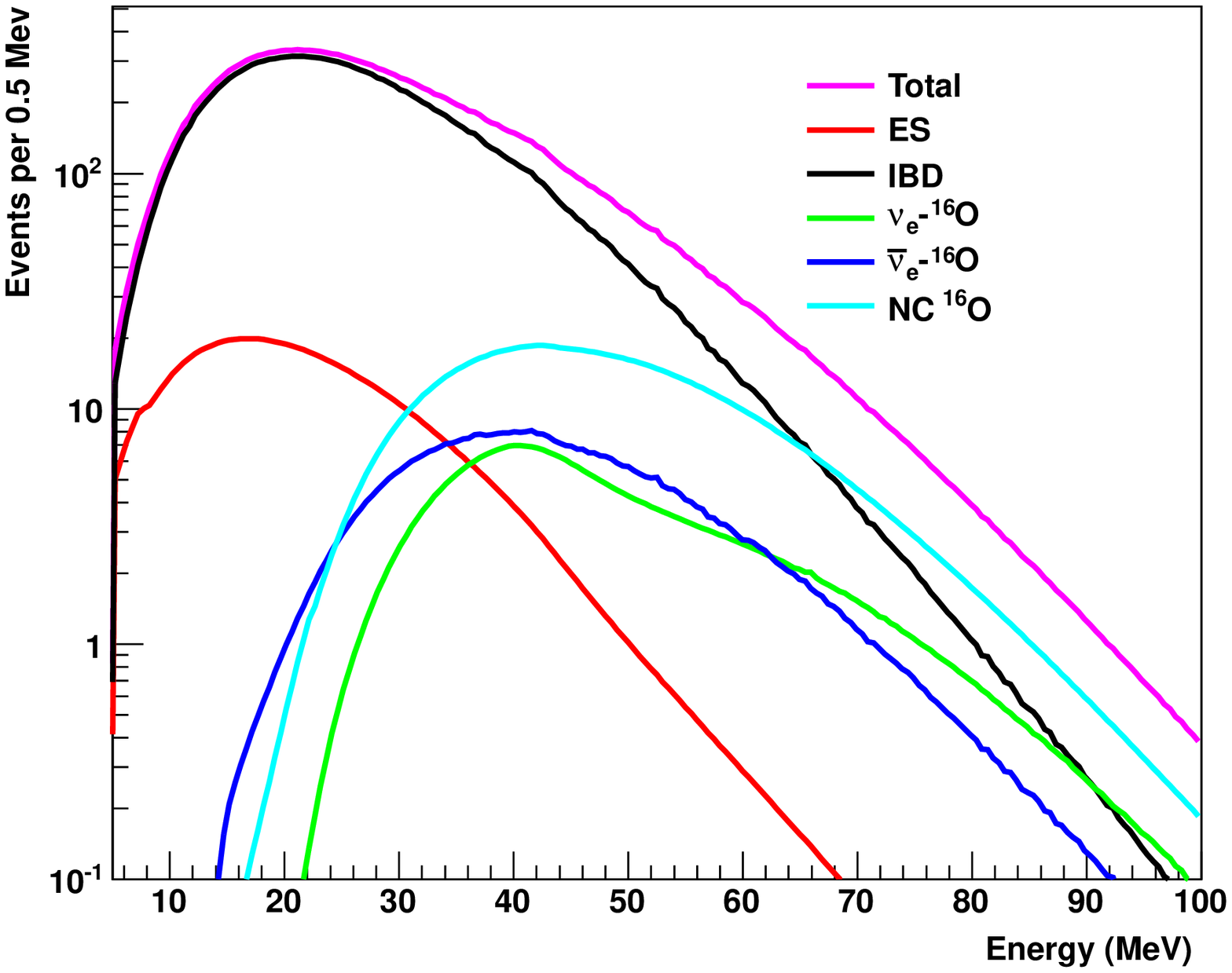,height=3in}}
\centerline{\psfig{figure=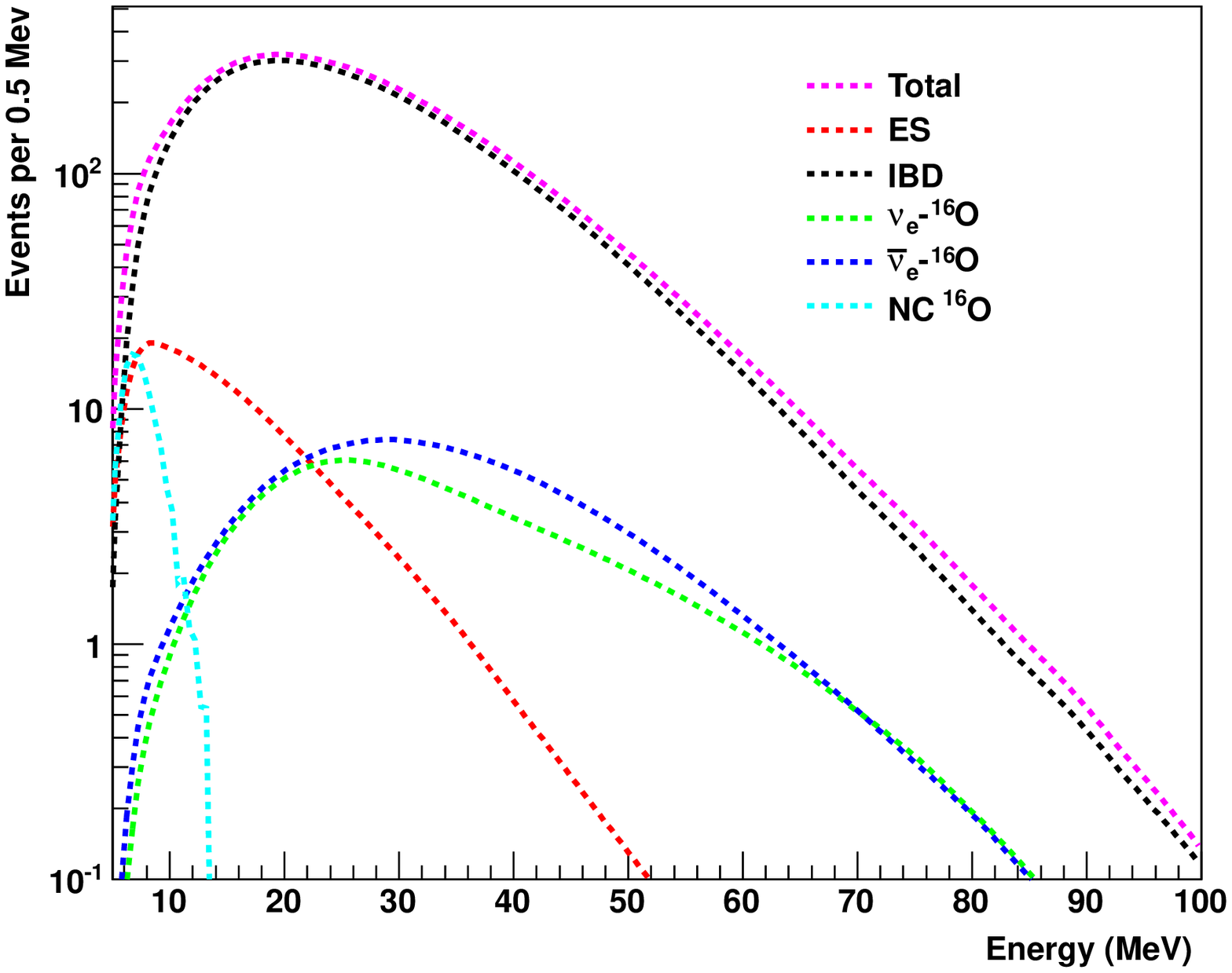,height=3in}}

  \caption{Event rates in 100~kt of water for the GKVM (Gava-Kneller-Volpe-McLaughlin) model and 30\% photomultiplier tube
    coverage (events per 0.5~MeV). Top: interaction rates as a
    function of true neutrino energy.  Bottom:  smeared
    rates as a function of detected energy.}
  \label{figure4}
\end{figure}

Figure~\ref{figure4} shows an example of expected event rates from a water Cherenkov detector, computed according to \texttt{http://www.phy.duke.edu/$\sim$schol/snowglobes}  and
assuming detector response similar to that of Super-K~I~\cite{Akiri:2011dv}.

\subsection{Long-String Water Cherenkov Detectors}

Long-string water Cherenkov detectors are arrays of long vertical strings of PMTs in water or ice.
Although such detectors are nominally designed for the study of very high energy astrophysical neutrinos (up to the TeV range or higher), such detectors may also be capable of supernova neutrino detection, assuming that the background rates are sufficiently low~\cite{Halzen:1994xe,Halzen:1995ex}.  The idea is that neutrino interactions in the water or ice surrounding the PMTs (mostly IBD) will create enough photons to produce a coincident increase in the single-count rate in the PMTs, observable over background counts.  The IceCube detector, which consists of 86 strings with 60 PMTs each and is embedded 1-2~km below the surface of the Antarctic ice, has had a supernova trigger installed~\cite{Abbasi:2011ss} and has demonstrated sensitivity to supernovae within the Milky Way. 
Future infill upgrades
of higher-PMT-density strings could improve supernova sensitivity by allowing for the detection of coincidences between PMTs from individual neutrino interactions.

Currently, IceCube cannot reconstruct individual neutrino interaction events, so it is insensitive to spectral and directional information.  However, thanks to the huge photon statistics, it has very good timing, which will render it sensitive to the time structure of the supernova burst (\textit{e.g.} References~\cite{Halzen:2009sm,Lund:2010kh}).
Some other long-string detectors, such as ANTARES~\cite{Ageron:2011pe}, are so noisy that they may not be able to trigger themselves but they may be able to receive external triggers to retain extra data.  The future KM3NET~\cite{Leisos:2012dk} detector may be able to suppress background by use of PMT coincidences.

\subsection{Liquid Argon Time Projection Chambers}

Liquid argon detectors will have excellent sensitivity to $\nu_e$ via the CC interaction on $^{40}$Ar,  
$\nu_e+{}^{40}{\rm Ar}\rightarrow e^{-}+{}^{40}{\rm K}^{*}$.
In principle, this is a taggable interaction,
for which the deexcitation $\gamma$s from $^{40}{\rm K}^{*}$  can be observed. 
The $\bar{\nu}_e$ interaction,
$\bar{\nu}_e+{}^{40}{\rm Ar}\rightarrow e^{-}+{}^{40}{\rm Cl}^{*}$ ,
will also occur and can be tagged via the pattern of $\gamma$s.
NC excitations are also possible, although little information is currently available in the literature about cross sections and observables.  Finally, there will be elastic scattering of neutrinos on electrons.

Large liquid argon detectors suitable for supernova neutrino detection are primarily large TPCs, in which ionization charge is drifted by an electric field and signals are collected on wire planes.  Using the time of arrival of charge at the readout planes, one can reconstruct three-dimensional track; particles can be identified by their rate of energy loss along a track.  Argon also scintillates, and scintillation photons collected by
PMTs enable fast timing of signals and enhance event localization within the detector.
Liquid argon TPC detection technology offers good energy resolution
and full particle reconstruction, unaffected by Cherenkov threshold. 
For sufficiently fine wire spacing, millimeter position resolution and very high quality tracking can be achieved.  Energy
thresholds as low as a few MeV may be possible.  
The direction of the scattered electron for elastic scattering can be determined~\cite{Bueno:2003ei}.  
Cosmogenic backgrounds should be a function of depth and location and are relatively unknown, although some preliminary studies exist~\cite{Barker:2012nb}.

A current example is ICARUS in Italy~\cite{Bueno:2003ei}, which will record events as long as suitable triggering can be implemented.
Future ideas include the LBNE liquid argon detector~\cite{Akiri:2011dv}, GLACIER~\cite{Angus:2010sz}  in Europe,
and possibilities in Japan~\cite{Hasegawa:2011zz}.  Smaller surface detectors such as MicroBooNE at Fermilab~\cite{Soderberg:2009rz} may also record some events.

\begin{figure}[htb]
  
\centerline{\psfig{figure=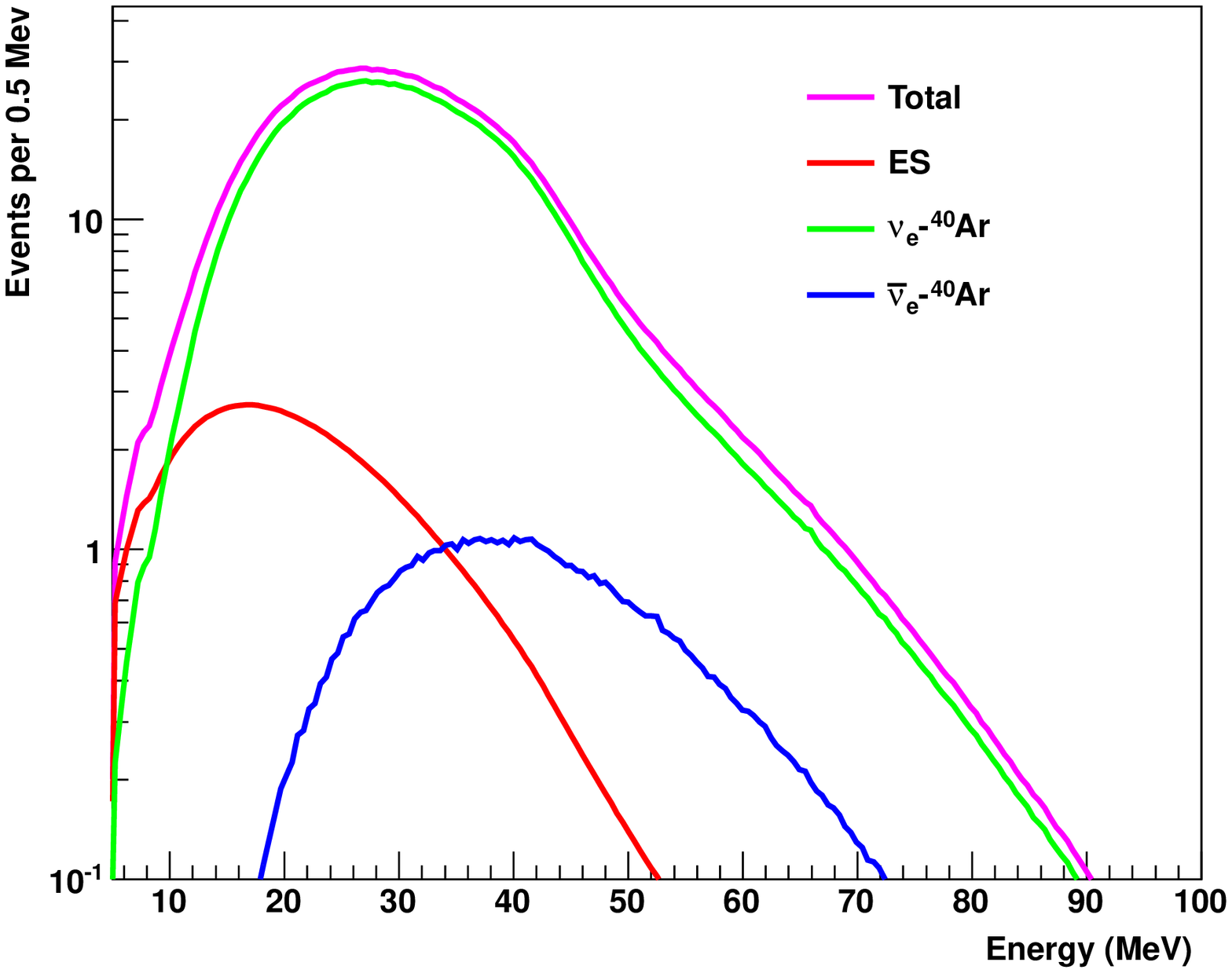,height=3in}}
\centerline{\psfig{figure=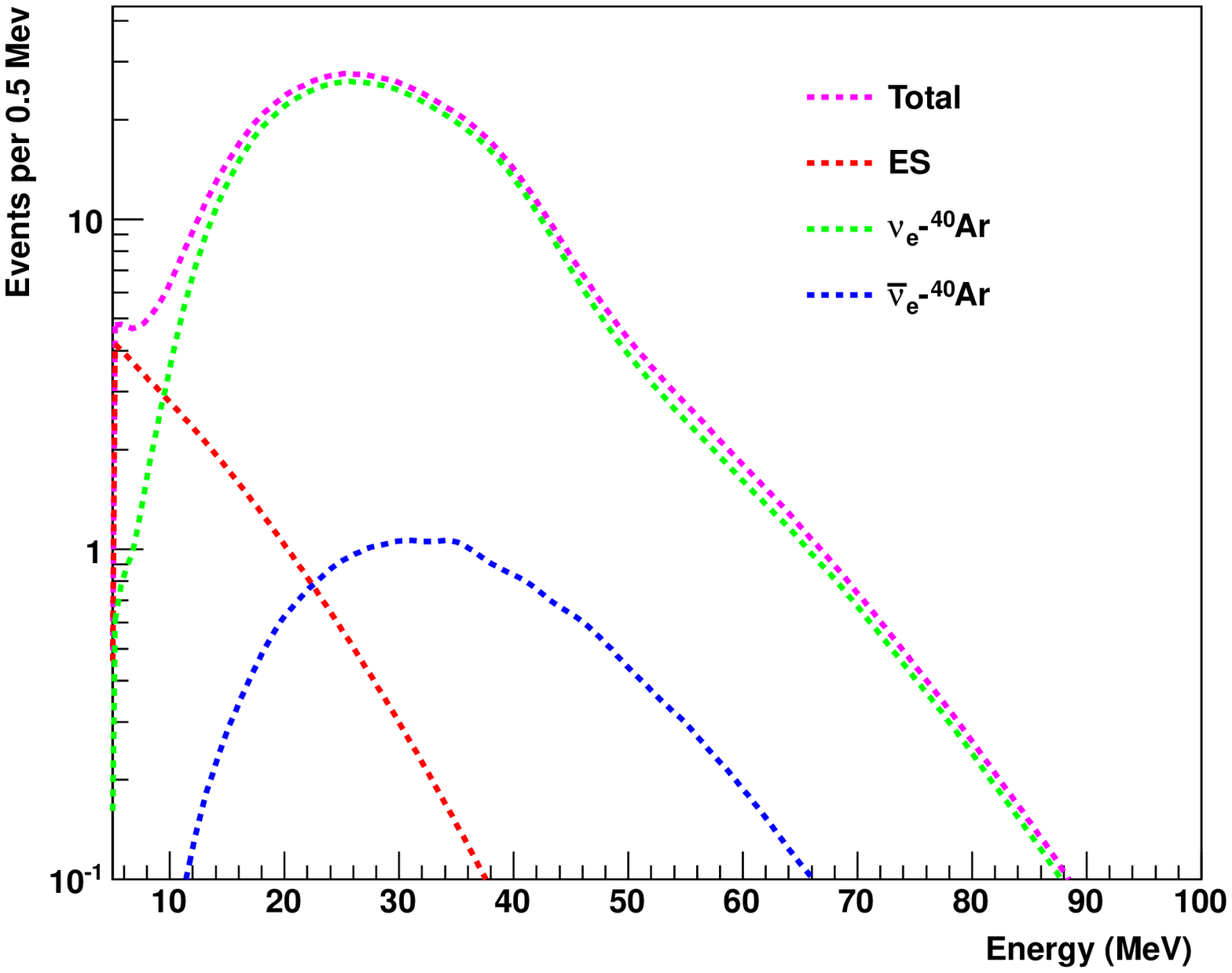,height=3in}}

  \caption{Event rates in 17~kt of argon for the GKVM (Gava-Kneller-Volpe-McLaughlin) model (events per 0.5~MeV). Top: interaction rates as a
    function of true neutrino energy. Bottom:  smeared
    rates as a function of detected energy.}
  \label{figure5}
\end{figure}

Figure~\ref{figure5} shows an example of expected event rates from an argon detector, computed according to \texttt{http://www.phy.duke.edu/$\sim$schol/snowglobes}  and
assuming the detector resolution from Reference~\cite{Amoruso:2003sw}.

\subsection{Detectors Employing Heavy Nuclei}\label{lead}

Interactions with
heavier nuclei such as iron and lead may yield quite high rates ~\cite{Hargrove:1996zv,Fuller:1998kb,Engel:2002hg,Boyd:2002cq,Zach:2002is} of both CC and NC
interactions.  Observables include leptons and ejected nucleons.  
Single- and multi-neutron ejections are possible.  Because the cross sections for single- and double-neutron ejections strongly depend on neutrino energy, a measurement of the relative rates of these processes provides information about the incoming neutrino spectra.
Lead is especially promising for neutron detection because of its stability and ease of handling in large quantities.  Furthermore, its most abundant isotopes have low neutron capture cross sections, making it nearly transparent to supernova-neutrino-induced neutrons, allowing the neutrons to reach a region of the detector system that is sensitive to neutron capture.

Although detectors employing iron and/or lead in
different configurations have been proposed (e.g., OMNIS~\cite{Boyd:2002cq,Elliott:2000su}),
the only current example of such a detector is HALO in Canada~\cite{Duba:2008zz}.  HALO 
detects neutrons by using recycled SNO $^3$He counters in combination with 79 tons
of lead.
The relevant interactions are
$\nu_e+{}^{A}{\rm Pb}\rightarrow e^{-}+{}^{A}{\rm Bi}^{*}$ 
and 
$\nu_x+{}^{A}{\rm Pb}\rightarrow \nu_x+{}^{A}{\rm Pb}^{*}$.
For both CC and NC cases, the resulting nuclei deexcite via nucleon emission.    
Antineutrino CC interactions are strongly suppressed by Pauli blocking because of lead's neutron excess.  
Although natural lead contains isotopes other than $^{208}$Pb, the neutrino cross section
for $A=208$ should be similar for other components~\cite{Kolbe:2000np,Engel:2002hg}.
In HALO, neutrons are moderated in polypropylene before being captured, and the energy loss of the particles resulting from neutron capture on $^3$He is recorded in proportional counters.  The detection efficiency of the capture of a neutron generated in the lead is on the order of tens of percent. There is no event-by-event energy information, but numbers of single- and double-neutron-producing interactions can be inferred and these relative numbers are very sensitive to the neutrino spectrum.
HALO2 is a proposed upgrade to approximately the kt scale.

\begin{figure}[htb]
 
\centerline{\psfig{figure=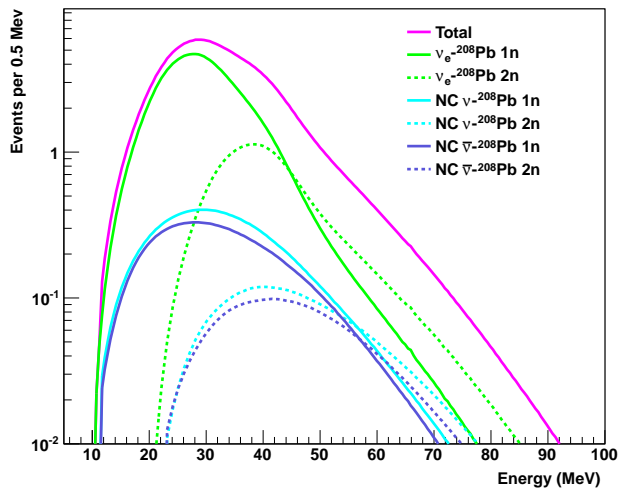,height=3in}}

 \caption{Event rates in 1~kt of lead (events per 0.5~MeV) for the GKVM (Gava-Kneller-Volpe-McLaughlin) model. The
   interaction rates are shown as a function of true neutrino energy.}
 \label{figure6}
\end{figure}

Reference~\cite{Vaananen:2011bf} explores the physics sensitivity for HALO.
Figure~\ref{figure6} shows an example of expected interactions in lead, computed according to \texttt{http://www.phy.duke.edu/$\sim$schol/snowglobes} .

\subsection{Low Threshold Detectors}\label{lowthresh}

Because of its very low energy threshold, only very clean, radioactively quiet detectors can detect nuclear recoils from coherent elastic neutrino-nucleus scattering; they must also have very good rejection of electronic recoils. 
Detectors developed for WIMP detection (or for neutrinoless double-$\beta$-decay searches or low-energy solar neutrino detection) fit the bill.  Examples are described in References~\cite{Gaitskell:2004gd} and~\cite{Akimov:2011zz}.   
Possibilities include
solid-state, noble liquid and bubble technologies, as well as new spherical TPC
proposals~\cite{Vergados:2006jg}.
Noble liquid detectors with xenon, argon or neon targets are particularly promising for scale-up to multiton scales; they have recoil energy thresholds of a few tens of keV.
These detectors are sensitive to all flavor components of the flux via the NC coherent interaction, and
some spectral information can be obtained via the measured recoil energy spectrum.
Expected supernova neutrino event rates are a few events per ton at 10~kpc, so
current-generation detectors will have only modest sensitivity.
However, future experiments
of 10-ton-or-greater scale are very promising, in that they may provide spectral information about all flavors.

\subsection{Summary of Detectors}

\begin{table}[h]

\caption{Summary of neutrino detectors with supernova sensitivity.
Neutrino event 
estimates are approximate for 10~kpc; note that there is significant variation by model.
Not included are smaller detectors
(e.g., reactor neutrino scintillator experiments) and detectors sensitive primarily to coherent elastic neutrino-nucleus scattering (e.g., WIMP dark matter search detectors).  The entries marked with an asterisk are surface or near-surface detectors and will have larger backgrounds. }
\label{tab:detectors}
\begin{tabular}{@{}cccccc@{}}%
\hline
Detector&Type &Mass (kt) &Location & Events & Live period\\ \hline

Baksan & C$_n$H$_{2n}$  & 0.33 & Caucasus & 50 & 1980-present\\ 
LVD & C$_n$H$_{2n}$& 1 & Italy& 300& 1992-present\\
Super-Kamiokande & H$_2$O& 32 & Japan& 7,000& 1996-present\\   
KamLAND & C$_n$H$_{2n}$& 1 & Japan& 300& 2002-present \\
MiniBooNE$^*$ &  C$_n$H$_{2n}$ & 0.7 & USA & 200 & 2002-present \\
Borexino& C$_n$H$_{2n}$& 0.3 & Italy& 100 & 2005-present \\ 
IceCube & Long string& 0.6/PMT & South Pole & N/A& 2007-present \\ 
Icarus &  Ar & 0.6 & Italy & 60 & Near future \\  
HALO &  Pb & 0.08 & Canada & 30 & Near future \\  
SNO+ & C$_n$H$_{2n}$& 0.8 & Canada& 300 & Near future \\    
MicroBooNE$^*$ &  Ar & 0.17 & USA & 17 & Near future \\ 
NO$\nu$A$^*$ &  C$_n$H$_{2n}$ & 15  & USA &  4,000& Near future \\
LBNE liquid argon &  Ar & 34 &USA  & 3,000  & Future \\
LBNE water Cherenkov &  H$_2$O & 200 &  USA& 44,000 & Proposed \\
MEMPHYS &   H$_2$O& 440 & Europe & 88,000 & Future \\
Hyper-Kamiokande &  H$_2$O & 540 & Japan & 110,000 & Future \\ 
LENA &   C$_n$H$_{2n}$& 50 &  Europe & 15,000 & Future \\ 
GLACIER &  Ar & 100 &  Europe& 9,000 & Future \\ \hline

\end{tabular}

\end{table}

Table~\ref{tab:detectors} summarizes existing and future detectors.  Note that the live detector mass used for supernova neutrino detection may be greater than the restricted ``fiducial'' mass often employed for physics analyses of typical beam, atmospheric, or solar neutrino experiments, because the background rates during a supernova burst are low compared with the signal rates.  For example, in Super-K, supernova burst events could be analyzed in the full 32-kt inner-detector volume, whereas the typical mass used for beam, atmospheric and solar neutrino analyses is 22.5~kt or smaller.

\section{POINTING TO THE SUPERNOVA WITH NEUTRINOS}\label{pointing}

It will be tremendously valuable to determine the direction of the supernova from the neutrino signal itself.  First, this information will be useful for an early alert.   For obvious reasons, astronomers want to know where to point their telescopes.  Second, a possible scenario is that the supernova will have no signal in electromagnetic radiation, or only a very faint signal, and pointing information will be useful for locating a remnant (or a progenitor from catalogs).  Doing so could, for example, allow one to learn the distance of travel by the neutrinos through the Earth for matter effect evaluation.

The most promising method for neutrino pointing is via elastic scattering (see Section~\ref{es}), in which the electron gets kicked in the direction of the neutrino.  In a Cherenkov detector, the directionality of the electron can be determined from the Cherenkov ring.  Because elastic scattering represents only a few percent of the total signal, the problem becomes one of finding a small bump on a nearly isotropic background.
In the absence of background, pointing quality goes roughly as $\sim 25^\circ/\sqrt{N}$, where $N$ is the observed number of elastic scattering events.
Reduction of the nearly isotropic background (mostly IBD) can be achieved through the use of neutron tagging with Gd (see Section~\ref{wc}).
When background is taken into account, the expected pointing quality for Super-K at 10~kpc~\cite{Beacom:1998fj,Tomas:2003xn} is about 8$^\circ$, which improves to approximately 3$^\circ$ with good IBD tagging.  For a megaton detector, better-than-1$^\circ$ pointing is achievable.

IBD positrons have only a mild anisotropy~\cite{Vogel:1999zy}, which nevertheless could be exploited in a very high statistics measurement.
In principle it is possible to recover some directional information in scintillator by reconstructing the relative positions of neutron capture and positron energy loss.  This IBD directionality has been measured in the CHOOZ scintillator detector for reactor $\bar{\nu}_e$ interactions~\cite{Apollonio:1999jg}, and future scintillator detectors may be able to exploit the effect.
Other possible methods include detection of TeV neutrinos from the supernova~\cite{Tomas:2003xn} (which point well, but may not accompany all core-collapse supernovae and may have significant delay), and use of the matter oscillation pattern in a detector with good energy resolution (which requires very large statistics)~\cite{Scholberg:2009jr}. 

In principle, determination of the supernova direction by use of triangulation can be achieved through relative timing of signals from multiple detectors around the Earth.  By using a time signal in two detectors, one can locate the supernova within a ring on the sky; with three time measurements, two intersecting rings narrow down the location to two spots; and with timing from four detectors, one can point to a single spot.
However triangulation is likely difficult in practice, at least on a short timescale.   Times of flight through the Earth are on the order of milliseconds; because the neutrino signal is spread over $\sim$10 s,
it requires large statistics~\cite{Beacom:1998fj}  or else a relatively high flux short-timescale component of the signal.
Nevertheless, the triangulation technique may be feasible for future megadetectors.
At this time and in the near future,  the best bet for supernova neutrino pointing is elastic scattering in a water Cherenkov detector.

\section{AN EARLY ALERT: SNEWS}\label{earlyalert}

Neutrinos emerge promptly from a collapsing star, on a $\sim$10-s timescale, and the first electromagnetic radiation may not
appear for hours or even days, depending on the nature of the stellar envelope.
Therefore, neutrinos can be used to provide an early warning of an imminent visible 
supernova.   

Very early light from extragalactic supernovae is rarely observed, so a prompt alert would give astronomers valuable time to find the supernova.
The environment immediately surrounding the progenitor
star would be probed by the initial stages of the supernova.  
Any effects of a close binary companion upon the blast would occur
very soon after shock breakout.  Advance warning could enable observation of ultraviolet and soft X-ray flashes, which are
predicted at very early times.  There could also be entirely
unexpected effects at early times.  A Galactic supernova is rare enough that it will be critical to save all available information.

The SN1987A neutrino events were recorded approximately 2.5~h before the inferred time of the supernova's first light; in real time, the experimentalists found the events in their data only after the fact~\cite{Koshiba:1992yb}.  The situation will be different for the next nearby supernova.
The SuperNova Early Warning System (SNEWS) is an international network of detectors that aims to provide an early alert to astronomers of a supernova's occurrence~\cite{Antonioli:2004zb}.   At the time of this writing, Super-K, 
LVD, IceCube, and Borexino are active participants in the network.  Each detector involved
sends a prompt datagram to a central computer if a signal consistent with a burst of supernova neutrinos is observed, and if there is coincidence within 10~s
an alert message is automatically sent.  The idea is to require a coincidence for a very high confidence early alert, so as to suppress false alarm background.
The criteria for the alert are fairly conservative; they require less than one
accidental coincidence per century.  The details of the coincidence conditions are described in References~\cite{Antonioli:2004zb} and \cite{Scholberg:2008fa}. 
Any pointing information sent to SNEWS will be forwarded to astronomers (although depending on the detectors involved, no pointing information may be initially available).

Note that some detectors that can provide
useful information are not necessarily capable of triggering themselves
on a supernova burst and may not be continuously archiving information.
They may be noisy and/or may not know what kind of signal to look for
from a supernova.  Some examples of detectors in this category include some of the long string detectors (ANTARES), gravitational-wave
detectors (if not all data are archived), and surface neutrino-sensitive
detectors with a high rate of cosmic-ray background.  A SNEWS neutrino
coincidence will be a high confidence indication that a supernova has
occurred.  Noisy supernova detectors would therefore arrange to use the
SNEWS coincidence as an input -- they would set up a buffering system to
record data (for hours or days, depending on the resources available) that
would routinely be overwritten but that could be saved to permanent
storage in the case of a SNEWS coincidence.  This approach would greatly
enrich the world's supernova data sample.

An interesting possibility for an early alert for the collapse itself is the detection of
neutrinos from a presupernova star~\cite{Odrzywolek:2003vn}.  Very large ($\sim$20 $M_\odot$) stars would produce pair-annihilation neutrinos during the final silicon-burning phase before core collapse.  The flux is small and the neutrinos are cool enough ($\sim$MeV average energies) to be difficult to detect, but in the few days before collapse there would be a distinctive hardening and an event rate increase.  Tagged IBDs from the high-energy tail of the spectrum are the most promising prospect, so scintillator or Gd-loaded water detectors would be required.  When  realistic backgrounds are accounted for, for 
a $\sim$100~kt  water detector the reach for this kind of early warning would probably be less than 1~kpc.  This represents only a small fraction of stars in the Milky Way, but it does include Betelgeuse, a nearby core-collapse candidate at 200~pc.

\section{EXTRAGALACTIC SUPERNOVA NEUTRINO SEARCHES}\label{extragalactic}

Table~\ref{tab:detectors} shows event rates at 10~kpc, which is just beyond the center of the Galaxy.  According to $1/D^2$ scaling, most detectors will observe robust bursts of events for any collapse that occurs within the Galaxy.   Beyond the edge of the Galaxy, there are few core-collapse candidate stars.
For a collapse in the LMC at 50~kpc, a few hundred events would be expected in Super-K, and approximately one event for Andromeda at 770~kpc.
A next-generation few-hundred-kt detector would detect dozens of events from Andromeda.  However, beyond that, the inverse square law strongly suppresses the signal and neutrino interactions become scarce.

Reference~\cite{Ando:2005ka} points out that with megaton-scale detectors, one might detect of the order of one neutrino per year from extragalactic sources.  The number of potential sources goes as $\sim D^3$.  There is evidence that the historical rate of supernovae within 10~Mpc or so is slightly higher than one would predict from simple scaling; one expects approximately one core collapse per year within that radius.  Furthermore, some black hole-forming collapses may produce hotter-than-usual spectra and therefore higher neutrino event rates, and they may
occur as often as once per decade within a few megaparsecs~\cite{Yang:2011xd}.

However, for a search for extragalactic supernova neutrinos, background suppression becomes critical.  The time window for an optical trigger would be several hours, and typical detectors would need to have a very low rate of background in this time window.  For example, at a 17-MeV threshold, Super-K's background rate is $\sim$1 per day~\cite{Ikeda:2007sa}; scaling to 1~Mt, this rate would be on the order of $\sim$10 background events in a several-hour time 
window, with only a $\sim$10\% chance of detecting a single signal event at 10~Mpc.  Requiring time clusters of two or more events reduces efficiency.
Backgrounds are typically highly site and detector dependent, however.  In 
particular, spallation backgrounds depend on the cosmic-ray rate and spectrum; therefore the
simple scaling from the rates in Reference~\cite{Ikeda:2007sa} may not necessarily be appropriate.
There are also plausible strategies to reduce backgrounds (see Section~\ref{diffuse}), 
and over megaton-decades, one might collect an observable signal over background. Reference~\cite{Kistler:2008us} explores what one might learn from  miniburst measurements of distant supernovae with a multimegaton detector.

Searches looking for coincidences with nonelectromagnetic collapse signatures, namely gravitational waves, may have more promise~\cite{Pagliaroli:2009qy,Leonor:2010yp} because the signal will also be prompt after core collapse.  However, a disadvantage of such searches is that the mechanism for gravitational-wave production and the nature of the signal are more uncertain; the distance range of gravitational-wave observability is poorly understood.


\section{THE DIFFUSE SUPERNOVA NEUTRINO BACKGROUND}\label{diffuse}

Looking even farther out for sources of neutrinos,  one can imagine measuring the flux of neutrinos from all the supernovae in cosmic history.  This so-called diffuse supernova neutrino background (DSNB) is sometimes referred to as the relic supernova neutrino flux.     
The physics of the DSNB is reviewed in References~\cite{Beacom:2010kk} and \cite{Lunardini:2010ab}. 

The DSNB flux depends on the historical rate of core collapse, average neutrino production, cosmological redshift effects and neutrino oscillation effects.
For neutrino energies above $\sim$19~MeV, estimates of the $\bar{\nu}_e$ component of the DSNB range from $\sim$0.1 to 1 cm$^{-2}$s$^{-1}$.

\begin{figure}[htb]
 
\centerline{\psfig{figure=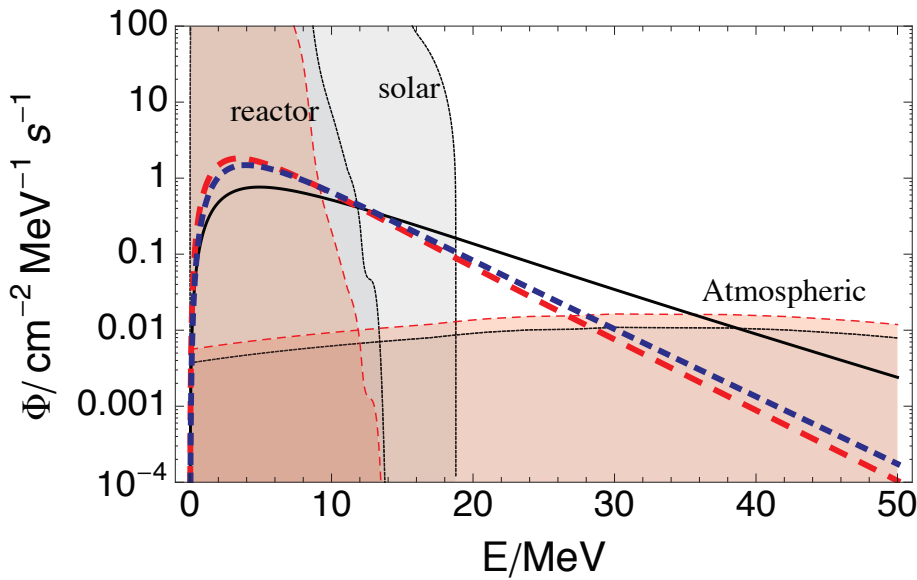,height=3in}}

 \caption{Diffuse supernova neutrino background (DSNB) $\bar{\nu}_e$  expected signal and backgrounds~\cite{Lunardini:2010ab}.  The background fluxes are shown for the Kamioka (solid, gray) and Homestake (dashed, red) sites for reactor and atmospheric neutrinos.  Solar neutrino background, which should be the same at both sites, is also shown. Three models for DSNB signal fluxes are shown. }
 \label{figure7}
\end{figure}

The detection interactions remain the same as for burst neutrinos;
however, the experimental issues become
entirely background dominated,
as there would be no external trigger at all and events would be measured singly. 
Overall, one would expect $\sim$0.1 IBD per kt per year in water or scintillator detectors from the $\bar{\nu}_e$ component of the DSNB. 
At lower energies, solar neutrinos dominate the neutrino background, although reactor $\bar{\nu}_e$ fluxes also contribute.  At higher energies, atmospheric neutrino backgrounds dominate. 
The atmospheric and reactor backgrounds vary by detector location. 
There is an energy window between $\sim$20 and 40 MeV in which the diffuse background dominates the neutrino flux incident on the Earth (Figure~\ref{figure7}) ~\cite{Lunardini:2010ab}.   
Detector-specific backgrounds for DSNB are the main issue in this energy window.   
In a water detector such as Super-K,
toward the lower end of this range, cosmic ray-induced spallation backgrounds leak into the signal window.  In the 20-50~MeV range,  atmospheric neutrino-induced muons dominate the background: In water, low-energy atmospheric CC $\nu_\mu$ can produce ``invisible'' muons below Cherenkov threshold, that then decay; the resulting Michel electrons mimic positrons from IBD.
So far, the best limits on the DSNB are from Super-K~\cite{Malek:2002ns, Bays:2011si}; the newer result has an updated limit of $< 2.8-3.0$ $\bar{\nu}_e$ cm$^{-2}$s$^{-1}$ for neutrino energies greater than $17.3~$MeV.  The new limit is slightly less stringent than the earlier result but is the outcome of improved analysis.  

The KamLAND scintillator
detector also has rather less stringent limits on the DSNB $\bar{\nu}_e$ flux,
but they extend to somewhat lower energies~\cite{Gando:2011jza}.
The background for DSNB $\bar{\nu}_e$ in scintillator differs from that in water:  The muons are not invisible because there is no Cherenkov threshold, and IBD interactions can be effectively tagged with neutrons.
However, NC interactions of atmospheric neutrinos on carbon (ejected neutrons in coincidence with deexcitation $\gamma$s) dominate the background in this energy range in scintillator.
The SNO heavy-water experiment~\cite{Aharmim:2006wq} has published a limit on the $\nu_e$ component of the DSNB flux, 70 cm$^{-2}$s$^{-1}$ at 90\% CL between 22.9~MeV and 36.9~MeV,  via a search for high-energy CC interactions of $\nu_e$ on deuterium.

Because tagging of $\bar{\nu}_e$ could reduce the Michel background (for atmospheric neutrino interactions unaccompanied by neutrons), the Gd proposal for Super-K~\cite{Beacom:2003nk,Kibayashi:2009ih} is promising for future measurements of the DSNB.  Future large water detectors are very promising, especially if Gd loaded~\cite{Akiri:2011dv},  as are future large scintillator detectors~\cite{Wurm:2011zn} as long as their NC background can be reduced.
Liquid argon detectors are sensitive to the $\nu_e$ component of the DSNB~\cite{Cocco:2004ac,Autiero:2007zj}.  For argon,  background is the key issue but is generally unknown at this time, although it could plausibly be small with good $\gamma$ tagging of $\nu_e$ absorption events and deep detectors.

The rate of DSNB neutrino events is steady, sure, and low: a constant rate of 0.1 events per kt per year can be expected.  In contrast, nearby collapse supernovae will provide copious bursts of neutrinos, but only sporadically: One expects thousands of neutrinos, but only every few decades, so building a detector with a short lifetime is a risky proposition due to potentially long Poissonian gaps between supernovae.  However, if one is prepared to wait, our own galaxy will provide, in the very long term, a much higher average rate of neutrino events: $\sim$10 events per kt per year.  The best experimental strategy is clearly a diversified portfolio: A large, long-running, low-background detector will win on both types of investment.

\section{PROSPECTS}\label{summary}

A Galactic supernova will offer unprecedented opportunities for diverse neutrino detectors around the world to gather critical information about astrophysics and  particle physics.   
Flavor sensitivity --not only the interaction rate but also the ability to tag different interaction channels-- will be critical for maximizing the science harvest from a burst observation. Current-generation detectors are sensitive primarily to $\bar{\nu}_e$; however, next-generation detectors will expand worldwide flavor sensitivity. 
Potential upgrades and new detectors will also enhance the prospects for detection of the so-far-unobserved DSNB.
If future megaton-scale detectors are built, prospects for a vast yield of information from a nearby burst are excellent, and we hope for a reach extending well beyond the Milky Way.

\section*{ACKNOWLEDGMENTS}
    The author's research activities 
are supported by the U.S. Department of Energy and the National Science
Foundation.  The author thanks her Super-K, LBNE, SNEWS and HALO collaborators (special thanks to Tarek Akiri and Cecilia Lunardini), as well
as the \textsf{SNOwGLoBES} contributors Farzan Beroz,
Rachel Carr,
Huaiyu Duan,
Alex Friedland,
Nicolas Kaiser,
Jim Kneller,
Alexander Moss,
Diane Reitzner,
David Webber, and
Roger Wendell.

\bibliography{refs}

\end{document}